\documentclass[11pt,twoside]{article}

\usepackage{asp2014}

\aspSuppressVolSlug
\resetcounters

\begin{document}

\title{SPARCL: SPectra Analysis and Retrievable Catalog Lab}

\author{St\'ephanie~Juneau,$^1$  Alice~Jacques$^1$, Steve~Pothier$^1$, Adam~S.~Bolton$^1$, Benjamin~A.~Weaver$^1$, Ragadeepika~Pucha$^2$, Sean~McManus$^1$, Robert~Nikutta$^1$ and Knut~Olsen$^1$}
\affil{$^1$NSF's NOIRLab, Tucson, AZ, United States; \email{stephanie.juneau@noirlab.edu}}
\affil{$^2$University of Utah, Salt Lake City, UT 84112, United States}

\paperauthor{St\'ephanie~Juneau}{stephanie.juneau@noirlab.edu}{0000-0002-0000-2394}{NSF's NOIRLab}{}{Tucson}{AZ}{85719}{United States}
\paperauthor{Alice~Jacques}{alice.jacques@noirlab.edu}{0000-0001-9631-831X}{NSF's NOIRLab}{}{Tucson}{AZ}{85719}{United States}
\paperauthor{Steve~Pothier}{steve.pothier@noirlab.edu}{}{NSF's NOIRLab}{}{Tucson}{AZ}{85719}{United States}
\paperauthor{Adam~S.~Bolton}{adam.bolton@noirlab.edu}{0000-0002-9836-603X}{NSF's NOIRLab}{}{Tucson}{AZ}{85719}{United States}
\paperauthor{Benjamin~A.~Weaver}{benjamin.weaver@noirlab.edu}{}{NSF's NOIRLab}{}{Tucson}{AZ}{85719}{United States}
\paperauthor{Ragadeepika~Pucha}{}{0000-0002-4940-3009}{University of Utah}{Department of Physics and Astronomy}{Salt Lake City}{UT}{84112}{United States}
\paperauthor{Sean~McManus}{sean.mcmanus@noirlab.edu}{}{NSF's NOIRLab}{}{Tucson}{AZ}{85719}{United States}
\paperauthor{Robert~Nikutta}{robert.nikutta@noirlab.edu}{}{NSF's NOIRLab}{}{Tucson}{AZ}{85719}{United States}
\paperauthor{Knut~A.~G.~Olsen}{knut.olsen@noirlab.edu}{0000-0002-7134-8296}{NSF's NOIRLab}{}{Tucson}{AZ}{85719}{United States}



\begin{abstract}
SPectra Analysis and Retrievable Catalog Lab 
(SPARCL) at NOIRLab's Astro Data Lab was created to efficiently serve large optical and infrared spectroscopic datasets. It consists of services, tools, example workflows and currently serves spectra of over 7.5 million stars, galaxies and quasars from the Sloan Digital Sky Survey (SDSS) and the Dark Energy Spectroscopic Instrument (DESI) survey. We aim to eventually support a broad range of spectroscopic datasets that will be hosted at NOIRLab and beyond. Major elements of SPARCL include capabilities to discover and query for spectra based on parameters of interest, a fast web service that delivers desired spectra either individually or in bulk as well as documentation and example Jupyter Notebooks to empower users in their research. 
More information is available on the SPARCL website\footnote{\url{https://astrosparcl.datalab.noirlab.edu}}.
\end{abstract}



\section{Introduction}

Wide area spectroscopic surveys have been producing a rapidly growing number of astronomical spectra. For instance, the Sloan Digital Sky Survey (SDSS) with the (extended) Baryonic Oscillation Sky Survey ((e)BOSS) have obtained spectra of over five million stars, galaxies and quasars over the course of about 20 years. More recently, the Dark Energy Spectroscopic Instrument (DESI) survey began a five-year campaign to obtain over 40 million spectra of galaxies and quasars as well as 10 million spectra of stars. We created SPectra Analysis and Retrievable Catalog Lab (SPARCL) to maximize the scientific return of such large spectroscopic datasets by providing users with a spectral database enhanced with a server and client for efficiently finding, accessing and retrieving the data of interest. For convenience, the client is installed at the NOIRLab Astro Data Lab\footnote{\url{https://datalab.noirlab.edu}} science platform \citep{Fitzpatrick+2014,Nikutta+2020} and can be instantiated directly from a notebook in the Data Lab JupyterLab environment \citep{Juneau+2021}. The client can also be installed locally by users. 
The SPARCL public API allows other clients and requests to interface with the data. After introducing SPARCL's data holdings and functionality, we highlight example use cases and briefly describe our future plans.


\section{Technical Overview}

SPARCL consists of the following main components:

\medskip
\noindent {\bf Database}: contains one-dimensional spectra and metadata. The focus is on optical spectra as a starting point (Table~1) with a plan to expand to infrared spectra. 

\medskip
\noindent {\bf Server}: provides web-service access to the database. The server can be accessed via the documented API\footnote{\url{https://astrosparcl.datalab.noirlab.edu/sparc/docs/}} using any language that supports HTTP Requests and Responses.

\medskip
\noindent {\bf Client}: enables Python-based data discovery and retrieval. The provided client simplifies and sometimes combines access to server web-services.

\medskip
\noindent {\bf Documentation}: separately describes the server API and the client package. For the server, users can try each API call directly from the documentation. In addition, documentation encompasses a User Manual and tutorial Jupyter notebooks, including a comprehensive {\it How to use SPARCL} notebook\footnote{\url{https://github.com/astro-datalab/notebooks-latest/blob/master/04_HowTos/SPARCL/How_to_use_SPARCL.ipynb}}.


\section{Data Holdings}

The first data sets ingested and served by SPARCL are public data releases from the large spectroscopic surveys SDSS and DESI as listed in Table~1. Those consist of one-dimensional spectra of stars, galaxies and quasars as well as data from redshift catalogs.

\begin{table}[!ht]
\caption{SPARCL spectroscopic data holdings}
\smallskip
\begin{center}
{\small
\begin{tabular}{llrrrr}  
\tableline
\noalign{\smallskip}
{\tt DATASETGROUP} & \texttt{DATA\_RELEASE} & N(spectra) & N(stars) & N(galaxies) & N(quasars) \\
\noalign{\smallskip}
\tableline
\noalign{\smallskip}
SDSS\_BOSS & SDSS-DR16 & 1,798,901 & 642,465 & 978,243 & 178,193 \\
SDSS\_BOSS & BOSS-DR16 & 3,918,000 & 531,731 & 2,212,769 & 1,173,500 \\
DESI       & DESI-EDR  & 2,044,588 & 640,647 & 1,303,237 & 100,704 \\
\noalign{\smallskip}
\tableline
\noalign{\smallskip}
All &  & 7,761,489 & 1,814,843 & 4,494,249 & 1,452,397 \\ 
\noalign{\smallskip}
\tableline\
\end{tabular}
}
\end{center}
\end{table}
\vspace{-10pt}

The spectra as well as the metadata are stored in a Postgres database. Each record corresponds to one spectrum and the fields are divided into three categories:
\begin{enumerate}
    \item {\bf CORE}: Basic parameters that are defined consistently for all spectra in the data-base, and that can be used for data discovery. 
    These are parameters such as right ascension (RA), declination (Dec), exposure time, instrument, etc.
    \item {\bf SPECTRA}: Vector fields for spectral flux data and related data such as wavelength and inverse variance. These are standardized across data sets as much as possible.
    \item {\bf AUX}: Additional auxiliary parameters associated with a spectrum that are not standardized across data sets. These cannot be used for data discovery but can be retrieved as part of the output for sample refinement and/or scientific analysis.
\end{enumerate}


Data ingestion requires the use of a {\tt personality} script to handle the peculiarities of each data set including the type of input files and their data model. Next, the data holdings will be augmented with the DESI DR1 dataset, which is a factor ten larger than DESI EDR with $\sim$20~million spectra. In the future, the data ingestion process will be streamlined so that other surveys or datasets can be incorporated into SPARCL. 

\section{User-Facing Functionality}

\noindent {\bf Data Discovery}: Can be done via SPARCL using the \texttt{client.find()} method or using SQL queries on the \texttt{sparcl.main} database table at the Astro Data Lab. A call to \texttt{client.find()} can include constraints on any of the CORE fields via the \texttt{constraints} argument, which accepts a dictionary with field names as keys and the constraint conditions as values. 
Depending on the field, the constraint conditions can either be a list or a range of desired values. The \texttt{client.find()} method returns a results-object which has an attribute called \texttt{ids} that is a list of the unique identifiers of all records in the SPARCL database that satisfy the supplied data-discovery conditions. This list of identifiers can then be used to retrieve spectra as described below.

\medskip
\noindent {\bf Data Access and Retrieval}: Given a list of unique spectrum identifiers found from 
data discovery or otherwise constructed, 
the \texttt{client.retrieve()} method provides access to the spectra themselves. Users can specify the fields to be included in the results of \texttt{client.retrieve()} via the \texttt{include} argument. Including fewer fields generally increases performance. The SPARCL Field Catalog page\footnote{\url{https://astrosparcl.datalab.noirlab.edu/sparc/sfc/}} indicates which fields are returned by default (only a minimal set), and also which fields are returned by specifying \texttt{include=’ALL’} (which includes all scientific fields, but excludes some internal SPARCL bookkeeping fields.) 

\section{Example Use Cases}

\noindent {\bf Spectral Stacking:} 
The Astro Data Lab is used to query SDSS magnitudes and create eight bins of $g-r$ color. In each bin, we retrieve spectra of a random subset of $N$ stars with SPARCL and combine them into a composite spectrum (Fig.~\ref{fig:stacks}). 
Using $N=10$ spectra per bin (80 total; left panel) takes $\sim$5 sec
but the composite spectra remain noisy and the trend with color is not strictly monotonic due to large variations among individual spectra. Using $N=500$ spectra per bin (total 4000 spectra; right panel) takes $\sim$90 sec and yields high signal-to-noise composite spectra with a clear trend with $g-r$ colors. This use case illustrates the performance of SPARCL for thousands of spectra\footnote{Also see notebook: \url{https://github.com/astro-datalab/notebooks-latest/blob/master/03_ScienceExamples/EmLineGalaxies/01_EmLineGalaxies_SpectraStack.ipynb}}.

\articlefiguretwo{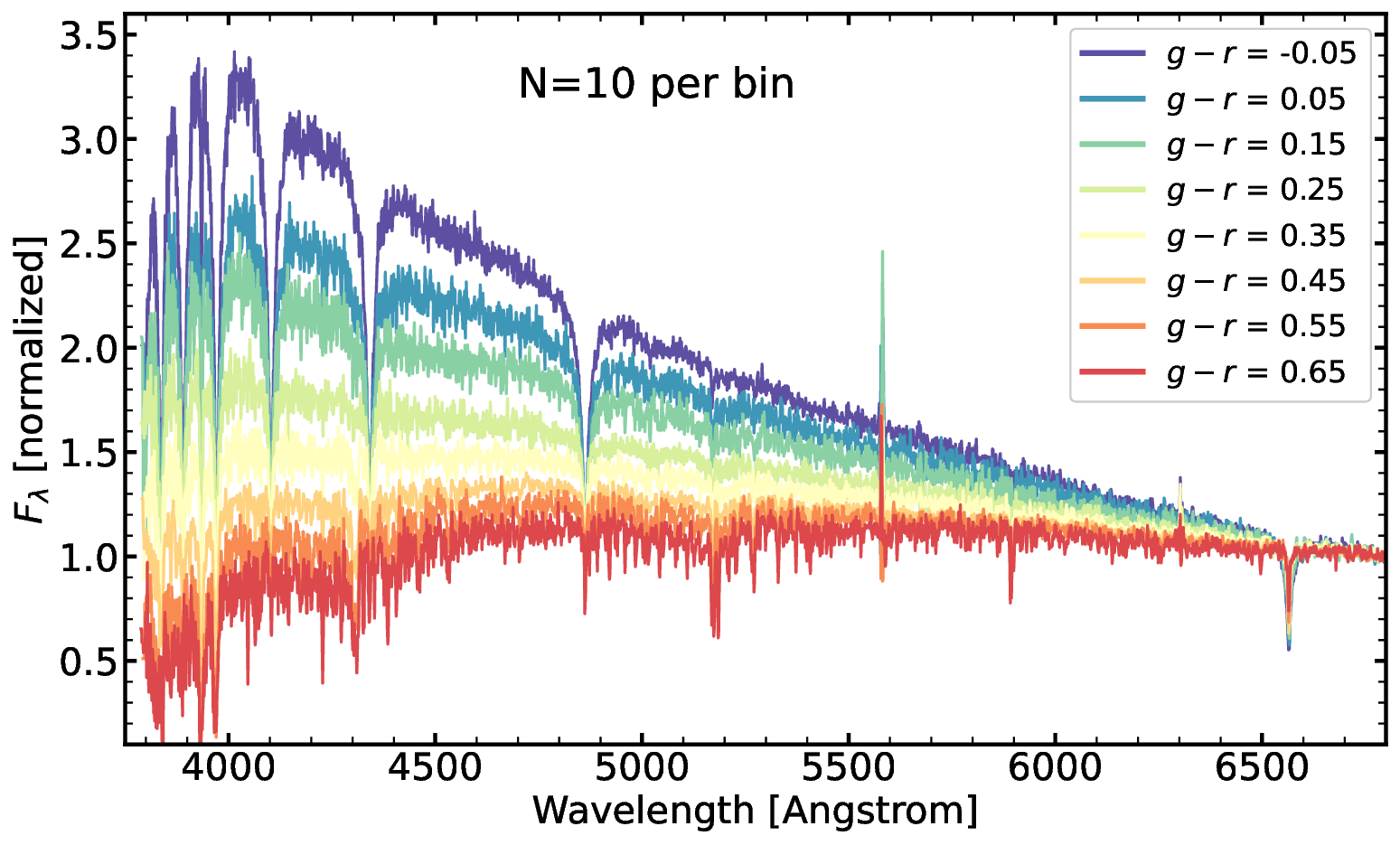}{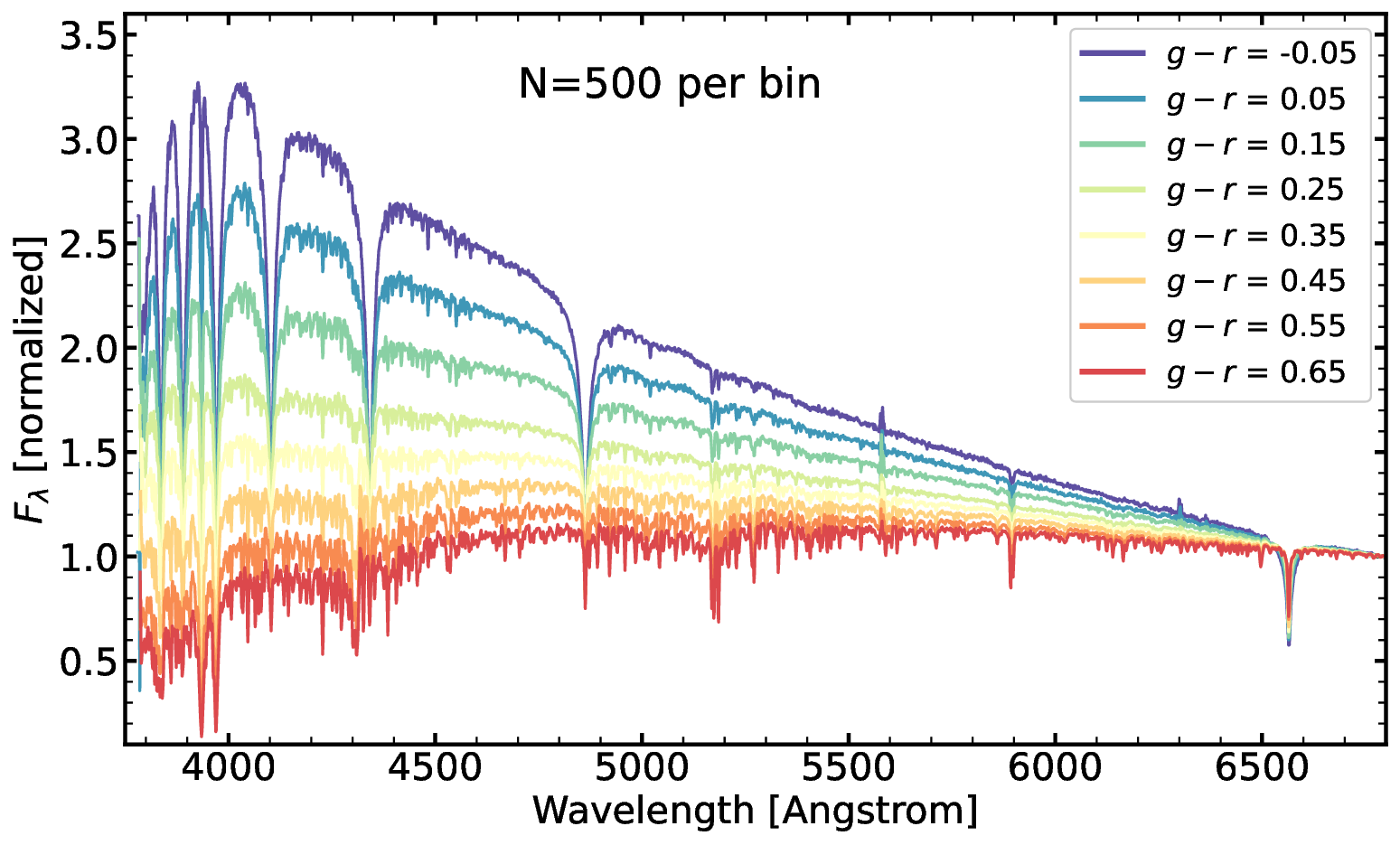}{fig:stacks}{Composite SDSS stellar spectra in bins of $g-r$ colors. Spectra are retrieved from SPARCL, stacked, and normalized at 6800~Angstrom. \emph{Left:} $N=10$ spectra per bin, which takes $\sim$5~seconds for the total of 80 spectra. \emph{Right:} $N=500$ spectra per bin, which takes $\sim$90~seconds for a total of 4000 spectra.}

\medskip
\noindent {\bf Interface with a Spectral Viewer:}
Spectra retrieved with SPARCL can be easily transformed into
inputs for a few spectrum visualization tools. Currently two
visualization tools are supported: 
Prospect\footnote{\url{https://github.com/desihub/prospect}} and Specviz\footnote{\url{https://jdaviz.readthedocs.io/en/latest/specviz/}}. 
In both cases, the visualization tools accept inputs as \texttt{specutils.Spectrum1D} 
objects\footnote{\url{https://specutils.readthedocs.io/en/stable/api/specutils.Spectrum1D.html}}.
The interface with spectral viewers is demonstrated in 
Astro Data Lab notebooks\footnote{\url{https://github.com/astro-datalab/notebooks-latest/blob/master/04_HowTos/SPARCL/} (e.g., \texttt{Plot\_SPARCL\_Spectra\_with\_Prospect.ipynb})}, which include examples for SDSS, BOSS and DESI data.
We expect that in the near future, the SPARCL client itself will provide the 
transformation to \texttt{specutils.Spectrum1D} objects,
which will make the interface even simpler.



\section{Summary and Future Plans}

SPARCL was developed to efficiently search and retrieve spectra from large surveys. It includes a database storing over 7.5 million optical spectra from SDSS and DESI, a server with a documented API, and a client that is installed at the Astro Data Lab and can be installed locally by users. Looking ahead, we will expand the data holdings to a broader variety of spectroscopic surveys. We will also develop enhanced functionality both server-side (e.g., aligning spectra) and client-side (e.g., output format conversion).

\acknowledgements SPARCL is part of the Community Science and Data Center (CSDC) program at NSF's NOIRLab, which is operated by the Association of Universities for Research in Astronomy, Inc. under a cooperative agreement with the NSF. 
Data acknowledgements are available on the survey websites for SDSS\footnote{\url{https://www.sdss4.org/collaboration/citing-sdss/}} and DESI\footnote{\url{https://data.desi.lbl.gov/doc/acknowledgments/}}.

\bibliography{P117}  


\end{document}